\documentclass[10pt,onecolumn,twoside]{IEEEtran}

\usepackage{amsmath,amsopn,amssymb}
\usepackage{graphicx,xspace,color}
\usepackage{epsfig,subfigure}
\usepackage{longt able,multirow}
\usepackage{array,float}
\usepackage{cite}
\usepackage[nolist]{acronym}
\usepackage{soul}
\usepackage{booktabs}

\begin{document}

\begin{acronym}
 \acro{BS}{base station}
\acro{SA}{slotted ALOHA}
\acro{IRSA}{irregular repetition slotted ALOHA}
\acro{SIC}{successive interference cancellation}
\acro{PLR}{packet loss rate}
\acro{MAC}{Medium access control}
\acro{BP}{belief propagation}
\acro{amacc}[A-MAC]{additive multiple access collision}
\acro{PMF}[p.m.f.]{probability mass function}
\acro{ic}[IC]{interference cancellation}
\acro{crdsa}[CRDSA]{contention resolution diversity slotted ALOHA}
\acro{iid}[iid]{independent and identically distributed}
\acro{mp}[MP]{message-passing}
\acro{ldpc}[LDPC]{low-density parity-check}
\acro{m2m}[M2M]{machine-to-machine}
\acro{iic}[IIC]{iterative interference cancellation}

\end{acronym}

 \title{ Finite Length Performance of  Random Slotted ALOHA   Strategies}
  \author{Konstantinos Dovelos,  Laura Toni,~\IEEEmembership{Member,~IEEE}, Pascal Frossard,~\IEEEmembership{Senior Member,~IEEE}
\begin{small}
\thanks{K. Dovelos is with  Aristotle University of Thessaloniki (AUTh), Greece. Email: \texttt{kdovelos@ece.auth.gr}. L. Toni, and P. Frossard are with \'Ecole Polytechnique F\'ed\'erale de Lausanne (EPFL), Signal Processing Laboratory - LTS4, CH-1015 Lausanne, Switzerland. Email: \texttt{\{laura.toni, pascal.frossard\}@epfl.ch.} 
}
\thanks{This work was partially funded by the Swiss National Science Foundation (SNSF) under the CHIST- ERA project CONCERT (A Context-Adaptive Content Ecosystem Under Uncertainty), project nr. FNS 20CH21 151569.}
\end{small}
}%
\maketitle
\thispagestyle{empty}

\begin{abstract}
  Multiple connected devices sharing  common wireless resources might create interference  if they   access  the channel simultaneously.  Medium access control (MAC) protocols generally  regulate the access of the devices  to the shared channel to limit signal interference.  In particular, irregular repetition slotted ALOHA (IRSA) techniques   can achieve  high-throughput performance when interference cancellation methods are adopted to recover from collisions. 
In this work, we   study   the finite length performance   for IRSA schemes by building on the analogy between successive interference cancellation and iterative belief-propagation on erasure channels.  We use a novel combinatorial derivation based on the matrix-occupancy theory  to compute the error probability  and we validate our method with simulation results.  
\end{abstract}

\section{Introduction}
 When  networked devices share common wireless resources, signal interference might be experience.  \acf{MAC} strategies need to  properly control users transmission to limit this interference~\cite{Gupta,Yih,Moulin}. However, in future networks a massive number of devices will be connected to the Internet  (e.g.,   Internet of Things and machine-to-machine communications) and MAC protocols need to be more and more distributed.   
Random \ac{SA}    with \ac{SIC} strategies, for example,  have recently gained   attention  because  they do not require  coordination, and   they are able to recover from interfering signals.

Bipartite graphs are a useful framework to study random MAC strategies or, more generally,   transmission of successive signals from several sources   in different time slots.   
When edges in the bipartite graph are randomly generated, the analysis of \ac{BP} decoding is usually performed asymptotically, i.e., for an infinite number of sources and time slots. Finite length analysis has been investigated when edges are randomly selected from the transmission time slots, as the case of   finite length analysis for LDPC codes \cite{richardson:B08}. However, the reverse case in which the source nodes randomly create the edges is still an open topic that we  address in this work. 

In this work, we consider   random \ac{SA}    with \ac{SIC} strategies as the main target  application, where   each source sends  information to a central \ac{BS} in time slots that are uniformly selected at random independently  from the other sources. Packets sent in the same time slot   from different users interfere among each others and cannot be immediately decoded. However, \ac{SIC} strategies are able to mitigate the effect of these collisions through  iterative  message-passing techniques and recover corrupted data at the decoder. Within this framework,  we study the decoding performance  of BP schemes in finite length settings, namely for small MAC frame size.  Within a MAC frame, each source follows a  transmission probability distribution that drives the replication rate of the sources, hence the performance of the system. Our objective   is to compute  the decoding error probability, i.e., the probability of not  decoding correctly the  source information.    We  first introduce a combinatorial derivation of the packet collision probability  using the matrix occupancy framework.  Then, we evaluate iteratively the decoding error probability by studying the number of collisions that can actually be resolved by interference cancellation. 
The proposed  analysis is exact but it has a computational complexity that grows   with the  MAC frame size. We therefore  show how achieve an approximated but still accurate analysis at a  reduced computational cost.     Simulation results validate   our   study  in different transmission  settings with small MAC frames.

 In the seminal work of \cite{Liva:J11},  a key connection has been drawn  between    SIC strategies in \ac{IRSA} and the  iterative \ac{BP} decoder of erasure codes on graphs.   This has opened the possibility to apply  theory of  rateless codes   to IRSA schemes and analyze their performance\cite{Paolini_Arxiv:J14,StePop:J13}, which is essential to optimize users' transmission strategy (e.g., transmission probability)~\cite{Toni:J15}.   These   works are mainly focused on deriving asymptotic system performance for large MAC size frames. They however cannot be easily applied in optimizing resource allocation strategies in actual IRSA schemes, as shown in~\cite{Toni:J15}. 
   To the best of our knowledge, only the works in \cite{Paolini:C15,IvanovBAP15a} investigated finite-length performance analysis for IRSA scheme. Both   look at the average stopping sets and derive an upper bound on the error probability in IRSA.  These bounds have low computational complexity but they are not necessarily tight for very small MAC frames. 
In our work, we  rather  derive a  semi-analytic  analysis for  finite length  \ac{IRSA} schemes, which permits to compute error probabilities exactly, even for small frames.


\begin{figure}[t]
\begin{center}
\includegraphics[width=.6\linewidth,  draft=false]{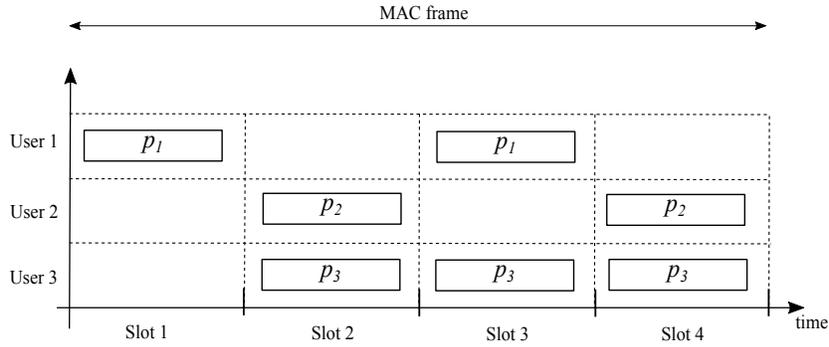}
\caption{Transmission example of \ac{IRSA} strategy with a MAC frame composed of four slots. Source $i$ sends  the source packet $p_i$. There are three users attempting a transmission according to the degree  vector $\mathbf{d} = [2, 2, 3]$.  }   \label{fig:matrixRepres_v2}
\end{center}
\end{figure}

\section{System Model}
\label{sec:system_model}
We consider a system of $k$  sources that communicate  with a common \ac{BS}. The  \ac{IRSA} strategy is the adopted MAC protocol~\cite{Liva:J11}. We assume   the time axis to be discretized in MAC frames,  each of those  composed of $t$ time  slots. Within a MAC frame, each  source  transmits $d$ replicas of the  source packet $p$,  as depicted in Fig. \ref{fig:matrixRepres_v2}.   The  $d$ distinct time slots used for transmission are  selected uniformly at random among the $t$ total available slots.    The replication rate $d$ is randomly selected by each user   following the transmission probability distribution  $\pmb{\Lambda}=[\Lambda_1, \ldots, \Lambda_{D_{max}}]$,  where $\Lambda_{d}$ is the probability that a user  transmits $d$   replicas, and $D_{max}$ is the maximum number of allowed packet replicas per MAC frame.   
Within a MAC frame,   each source  selects its replication rate  independently from the others, leading to  replication vector (named in the following as source degree vector) $\mathbf{d} = [d_1, \dots, d_k], \: d_i\in\{1,\dots, D_{max}\}$ that is experienced with    probability   $P_{\pmb{\Lambda}}(\mathbf{d}) = \prod_{i=1}^k P_{\pmb{\Lambda}}(d_i) = \prod_{i=1}^k \Lambda_{d_i}$. 

Each   realization of $k$ sources accessing the  time slots of a MAC frame   can be described by a $k\times t$ binary matrix $\mathbf{M} = (m_{ij})$, called \emph{collision matrix}, with rows and columns corresponding to users and slots, respectively.  We have    $m_{ij}=1$ if the $i$th user transmits in the $j$th slot,  and  $m_{ij}=0$ otherwise.  The collision matrix $\mathbf{M}$ associated with the example in   Fig. \ref{fig:matrixRepres_v2} is given by  
\\
\[ \mathbf{M} = 
\begin{pmatrix}
1 & 0 & 1 & 0 \\
0 & 1 & 0 & 1 \\
0 & 1 & 1 & 1 
\end{pmatrix}.
\]
The weight of a column $\mathbf{m}_j$ in $\mathbf{M}$ is given by   $\sum_{i=1}^{k}m_{ij}$ and it   represents  the number of  packets  sent in the time slot $j$. Thus, columns with unity weight, e.g., $[1   0 0]^T$, represent singleton slots that allow  an immediate decoding of the message. On the contrary,   columns with a weight greater than one, e.g., $[1 1 0]^T$, represent   slots in which messages collide and cannot be directly decoded.    Collided messages can however be recovered by \ac{SIC} strategies. If  packets are sent by two users in the same time slot  but one of them can be recovered from a  singleton slot,   then  the second packet can be decoded  by interference cancelation. For example, message $p_1$ in $\mathbf{M}$ is recovered from the first slot, which is a singleton one. Then, canceling the message $p_1$ from the other interfering messages we obtain $ \mathbf{M}^{\prime} = [ 0 \, 0 \, 0 \, 0;  0 \, 1 \, 0 \, 1;  0 \, 1 \, 1 \, 1 ]$ 
and message $p_3$ can also be decoded. 
As long as one singleton slot is experienced, the iterative decoding process proceeds. If the SIC process resolves all collisions, then no source packets are lost within the MAC frame of interest. If the SIC process   stops before completion, it leaves packets undecoded and the SIC process fails. 

In this work, we  are interested   in evaluating the probability of failure in the SIC process, i.e.,  the  probability that a packet is lost when transmitted through the IRSA protocol.  We denote this    \ac{PLR}  by     $P_L$,  
  and it  can be written as\footnote{For the sake of notation, we omit the dependency of the packet loss probabilities on $(k,t)$.}
\begin{equation}\label{eq:plr1}
P_L  = \sum_{u=2}^k \: \frac{u}{k} \: P_{\pmb{\Lambda}}(u) 
\end{equation}
where $P_{\pmb{\Lambda}}(u)$ is the probability of having $u$ unrecovered packets when  $k$ users transmit over a frame of $t$ slots with degree distribution $\pmb{\Lambda}$.  
We condition to a given degree distribution vector as follows
\begin{align}\label{eq:plr2}
P_L &= \sum_{\mathbf{d}\in\mathcal{D}} \: \sum_{u=2}^k \: \frac{u}{k} \: P_{\pmb{\Lambda}}(u\mid  \mathbf{d}) \: P(\mathbf{d}) \nonumber \\
&= \sum_{\mathbf{d}\in\mathcal{D}} \: \sum_{u=2}^k \: \frac{u}{k} \: P(u\mid  \mathbf{d}) \: \prod_{i=1}^k \Lambda_{d_i}
\end{align}
with $\mathcal{D}$ denoting the set of all the possible packet repetition vectors allowed by the distribution $\pmb{\Lambda}$. Denoting by $D$ the number of possible replication rates, i.e., replication rates with ${\Lambda}_{d}  > 0$,     $|\mathcal{D}| = D^k$. 
In the next section, we compute the    \ac{PLR}      $P_L$ for small    MAC frame size $t$. 
 

\section{Finite Length Performance} 
\label{sec:analsys} 

\subsection{  Matrix-Based Formulation} 
\label{sec:occupancy_matrix}
Because of the source independence,  collision matrices are equivalent in terms of  PLR  upon permutations (both across rows or columns). We can therefore study the IRSA performance by   only looking at the column vectors which are present within a given matrix $\mathbf{M}$. This is possible exploiting the combinatorial matrix-occupancy theory~\cite{Matrix_occ}, dealing with   sets of balls   randomly assigned into  groups of bins.  Random access channel problems can be viewed as   occupancy problems by considering packets and slots as balls and bins, respectively. {The number of bins with only one ball, for example, represents the number of  singleton slots. }  

In more details, let  ${\mathcal{C}}  = \{{\mathbf{c}_1}, {\mathbf{c}_2}, \ldots , {\mathbf{c}_{|\mathcal{C}|}} \}$  be the set of all possible column vectors that can be present in $\mathbf{M}$, with column $\mathbf{c}_q=[c_q^{(1)}, c_q^{(2)}, \ldots, c_q^{(k)}]^T$ taking values in $\{0,1\}^k$.
 Let us then define the \emph{occupancy vector}  $\mathbf{n}  = [n_{\mathbf{c}_1}, n_{\mathbf{c}_2}, \ldots , n_{\mathbf{c}_{|\mathcal{C}|}} ]$ associated with a matrix   $\mathbf{M}$  as a vector that shows how many times  each  column  in ${\mathcal{C}}$  is present in $\mathbf{M}$. Note that for the sake of notation, we omit the dependency of $\mathbf{n} $ from $\mathcal{C}$.
 Specifically, $n_{\mathbf{c}_q}$ is the number of times the column $\mathbf{c}_q$ is present in the matrix of interest.  For example, defining ${\mathbf{c}_1} = [1 \ 0 \ 0]^T, {\mathbf{c}_2} = [0 \ 1 \ 1]^T,$ and ${\mathbf{c}_3} = [1 \ 0 \ 1]^T$, the occupancy vector associated with $\mathbf{M}$  is  
$$\mathbf{n}  =\left[n_{\mathbf{c}_1}=1,\: n_{\mathbf{c}_2}=2, \: n_{\mathbf{c}_3}=1,\: n_{\mathbf{c}_{q, q>3}}=0  \right].$$  
Finally,  we define    ${\mathcal{C}}_l\subseteq {\mathcal{C}}$ as the subset of column  vectors with weight $w(\mathbf{c}_q) = \sum_{j=1}^{k}c_q^{(j)}= l$, and   $\mathcal{C}_{l,i} \subseteq {\mathcal{C}}_l $ as the subset of column vectors with  weight $l$ and $c_i = 0$.  
It is worth noting that each occupancy vector corresponds  to multiple collision matrices that are equivalent in terms of   PLR.

We are now interested in finding 
conditions under which an occupancy vector   represents a collision matrix in the case of $k$ sources, $t$ time slots, and degree vector $\mathbf{d}$. 
First,  we impose that  exactly  $t$ columns are present in the matrix:
\begin{align}\label{eq:constraint}
\sum_{q: \mathbf{c_q}\in \mathcal{C} } n_{\mathbf{c}_q}=t
\end{align} 
Then, we impose that the degree vector is respected. This means that an occupancy vector is feasible if it leads to a matrix in which exactly $d_i$ entries are non-zero in the $i$th row of the collision matrix. 
This translates in the following set of constraints   
\begin{align}\label{eq:constraint2}
\sum_{l=1}^{D_{max}} \sum\limits_{q: \mathbf{c_q}\in \mathcal{C}_{l,i}} \: n_{\mathbf{c}_q}  = t-d_i,\quad i=1,\dots, k .
\end{align} 
Since $\mathcal{C}_k$ has only one column vector (i.e., the vector with all $1$ entries) and $\mathcal{C}_{k-1}$ has $k$ possible column vectors (i.e., each vector with only one out of $k$ null entry),  we can impose the above $k+1$ constraints --- \eqref{eq:constraint} and  \eqref{eq:constraint2} --- by properly evaluating the occupancy of the $k+1$ column vectors   in $\mathcal{C}_k$ and $\mathcal{C}_{k-1}$.   Let us denote by   $\hat{\mathbf{n}}$   the reduced occupancy vector, defined as the  column vectors with weight at most $k-2$.  Formally,  $\hat{\mathbf{n}}=[n_{\mathbf{c_q}}]_{\mathbf{c_q}\in \hat{\mathcal{C}}}$, with $\hat{\mathcal{C}} = \mathcal{C} \setminus   \mathcal{C}_{k-1} \cup \mathcal{C}_k  $. 
We can then  decompose any occupancy vector as  $\mathbf{n}  = [ \hat{\mathbf{n}}  \ \mathbf{f}(\hat{\mathbf{n}}, \mathbf{d})]$, with  ${f}(\hat{\mathbf{n}}, \mathbf{d})$ representing  the occupancy  of the $k+1$ column vectors   in $\mathcal{C}_k$ and $\mathcal{C}_{k-1}$. These $k+1$  unknown  $\mathbf{f}(\hat{\mathbf{n}}, \mathbf{d})=[f_1(\hat{\mathbf{n}}, \mathbf{d}), \ldots, f_{k+1}(\hat{\mathbf{n}}, \mathbf{d})]$  are derived by imposing the constraints  \eqref{eq:constraint} and  \eqref{eq:constraint2}. 
If $ f_{i}(\hat{\mathbf{n}}, \mathbf{d})\geq 0, \forall i$, then the   occupancy vector   $[\hat{\mathbf{n}}  \ \mathbf{f}(\hat{\mathbf{n}}, \mathbf{d})]$ is a feasible one for the transmission settings  $(k,t,\mathbf{d})$.  We   define  $\mathcal{I}(\hat{\mathbf{n}})$   an indicator function such that  $\mathcal{I}(\hat{\mathbf{n}})=1$ if    $[\hat{\mathbf{n}}  \ \mathbf{f}(\hat{\mathbf{n}}, \mathbf{d})]$ is a feasible one for the transmission settings  $(k,t,\mathbf{d})$, and $\mathcal{I}(\hat{\mathbf{n}})=0$, otherwise.

\subsection{Packet Loss Probability}
Equipped with the matrix-occupancy representation, we can express  the   error probability $P(u \mid   \mathbf{d})$ in \eqref{eq:plr2}  as 
\begin{align} \label{eq:error_prob1}
 P(u \mid   \mathbf{d})    =  \sum\limits_{\hat{\mathbf{n}} } Q_u(k,  [\hat{\mathbf{n}}  \ \mathbf{f}(\hat{\mathbf{n}}, \mathbf{d})])  \: P(\hat{\mathbf{n}}  \mid  \mathbf{d})  \nonumber
\end{align}
where $P(\hat{\mathbf{n}}  |  \mathbf{d})$ is the  probability of experiencing an occupancy vector   $[\hat{\mathbf{n}}  \ \mathbf{f}(\hat{\mathbf{n}}, \mathbf{d})]  $, when $k$ users transmit over   $t$ slots  given    the repetition vector $\mathbf{d}$. The indicator function  $Q_u(k,  \mathbf{n})$ returns $1$ if the the SIC process with a collision matrix associated with $\mathbf{n} $ stops  at  $u$ undecoded packets and returns $0$ otherwise. We compute both terms below.

The probability $P(\hat{\mathbf{n}}  |  \mathbf{d})$   is zero if    $\mathcal{I}(\hat{\mathbf{n}})=0$, otherwise it is evaluated  as the ratio between the number of collision matrices with occupancy vector $[\hat{\mathbf{n}}  \ \mathbf{f}(\hat{\mathbf{n}}, \mathbf{d})]$ and the total number of collision matrices in the same transmission settings. The former   is given by the following multinomial coefficients
$$
 \frac{t!}{\prod\limits_{\mathbf{c_q}\in \hat{\mathcal{C}}}\: n_{\mathbf{c}_q}! \prod\limits_{i=1}^{k+1} f_i(\hat{\mathbf{n}}, \mathbf{d})!}  
$$
while the total number of collision matrices that can be experienced under the settings  $(k,t,\mathbf{d})$ is $  \prod_{j=1}^{k} \: \binom{t}{d_i}$ from the independency of the sources. 
This leads to 
\begin{equation}\label{eq:moc3}  
P(\hat{\mathbf{n}}  |  \mathbf{d}) =
 \begin{cases}
 \left[  \prod_{j=1}^{k} \: \binom{t}{d_i}\right]^{-1}   \frac{t!}{\prod\limits_{\mathbf{c_q}\in \hat{\mathcal{C}} }\: n_{\mathbf{c}_q}! \prod\limits_{i=1}^{k+1} f_i(\hat{\mathbf{n}}, \mathbf{d})!}       
  & \text{if }\mathcal{I}(\hat{\mathbf{n}})=1 \\
 0, & \text{otherwise}  
 \end{cases}
\end{equation}

%
%
 
We  then derive    $Q_u(k,  \mathbf{n})$   iteratively. We consider the $j$th iteration of the decoding process, where $k-j$   packets are undecoded,    and    $\mathbf{n}^{(j)}= [n_{\mathbf{c}_1}^{(j)}, n_{\mathbf{c}_2}^{(j)}, \ldots ]$ is  the occupancy vector of the  collision matrix at the $j$th decoding step.  Note that  $\mathbf{n}=\mathbf{n}^{(0)}$ is the occupancy vector  before the decoding process starts. At the $j$th iteration of the decoding process, one message is decoded only if there exists at least one weight-$1$ column vector, i.e., if $\exists \:{\mathbf{c}} \in \mathcal{C}_1  \text{ s.t. } n_{\mathbf{c}}^{(j)} > 0$.    

If the condition is satisfied, then the decoder can proceed to the next step.   At the decoding iteration $j+1$, there are $k -j  - 1$ undecoded  packets and  the occupancy vector of the collision matrix is denoted by  $\mathbf{n}^{(j+1)}$.  
The latter  is derived recursively from $\mathbf{n}^{(j)}$. Let us consider the column   vector with the $m$-th entry being non-zero, i.e., $\mathbf{c} \in \mathcal{C} \setminus  \cup_l  \mathcal{C}_{l,m}$, and let us denote by $\overline{\mathbf{c}}^{(m)}$ its complementary in $m$
 a column vector equal to $\mathbf{c}$ but  with the $m$-th entry set to zero. 
For example, if  $\mathbf{c}= [1 1 0 0 1]$, then $\overline{\mathbf{c}}^{(2)}= [1 0 0 0 1]$. Then, in the case in which the $m$-th element of $\mathcal{C}_1$ has   $n_{\mathbf{c}_m}^{(j)} > 0$,  $\mathbf{n}^{(j+1)}$  can be written from  $\mathbf{n}^{(j)}$   as follows
\setcounter{equation}{5}
\begin{align} 
n_{\overline{\mathbf{c}}^{(m)}}^{(j+1)} &= n_{{\mathbf{c}}}^{(j)} + n_{\overline{\mathbf{c}}^{(m)}}^{(j)}   & \forall  \mathbf{c} \in \mathcal{C} \setminus  \cup_l  \mathcal{C}_{l,m}
 \nonumber \\
n_{\overline{\mathbf{c}}^{(m)}}^{(j+1)} &=0\
 \nonumber \\
 n_{{\mathbf{c}}}^{(j+1)} &=  n_{{\mathbf{c}}}^{(j)}   & \forall  \mathbf{c} \in  \{ \cup_l \mathcal{C}_{l,m} \setminus\overline{\mathbf{c}}^{(m)} \}
\end{align} 
We thus  recursively evaluate the indicator function $Q_u$ as   
\begin{align} 
Q_u(k - j,  \mathbf{n}^{(j)}) = Q_u(k - j -1, u, \mathbf{n}^{(j+1)}).
\end{align} 
If there are no weight-one columns in the collision matrix, the decoder terminates at iteration $j$ with $k-j$ undecoded packets and  $Q_u$ becomes 
\begin{equation}\label{eq:req1}
Q_u(k - j,  \mathbf{n}^{(j)}) = \left\{ 
\begin{matrix}
 1,  			\quad & k-j=u \\ \\
0,                                                                                       \quad &\textrm{otherwise}
\end{matrix} \right.
\end{equation}

Finally,   denoting by $\mathcal{N}$ the set of reduced occupancy vectors  $\hat{\mathbf{n}}$ such that  $\mathcal{I}(\hat{\mathbf{n}})=1$,  the decoding error probability of  \eqref{eq:plr2} results in 
 \begin{equation} \label{eq:plr2_B}  
P_L = \sum_{\mathbf{d}\in\mathcal{D}} \: \sum_{u=2}^k \: \frac{u}{k} \:  \sum\limits_{\hat{\mathbf{n}}  \in \mathcal{N}} Q_u(k,   [\hat{\mathbf{n}}  \ \mathbf{f}(\hat{\mathbf{n}}, \mathbf{d})]^{(0)})  \:      \frac{t!}{\prod\limits_{\mathbf{c_q}\in \hat{\mathcal{C}}  }\: n_{\mathbf{c}_q}! \prod\limits_{i=1}^{k+1} f_i(\hat{\mathbf{n}}, \mathbf{d})!}      \: \prod_{i=1}^k \frac{ \Lambda_{d_i}}{  \binom{t}{d_i}}
 \end{equation}

We now comment on the complexity of the proposed   semi-analytical study.  Both the combinatorial and iterative steps in   \eqref{eq:plr2_B} are performed over all possible degree vectors $\mathbf{d}\in\mathcal{D}$ and all possible reduced occupancy vectors $\hat{\mathbf{n}}  \in \mathcal{N}$. 
The cardinality   of $\mathcal{D}$ and $\mathcal{N}$ is given respectively by 
 $$|\mathcal{D}|=D^k \ \text{  and   }  \ |\mathcal{N}| \leq   \binom{\hat{\mathcal{C}}+t-1}{t}$$
with $\hat{\mathcal{C}}=\sum_{n=0}^{k-2} \binom{t}{n}$. The upper bond on $|\mathcal{N}|$ is derived as follows.   We first recall that $\hat{\mathcal{C}}$ is the  dimension of  the reduced occupancy vector $\hat{\mathbf{n}}$ and that the entries of $\hat{\mathbf{n}}$ need  to satisfy  \eqref{eq:constraint}. Looking at the problem as $t$ balls into $R$ bins,   the number of possible combinations of the reduced vector is  $\binom{\hat{\mathcal{C}}+t-1}{t}$. Among these, only the reduced occupancy vectors that satisfy \eqref{eq:constraint2}  belong  to $\mathcal{N}$.  

\begin{figure}[t]
\begin{center}
\includegraphics[width=.55\linewidth,  draft=false]{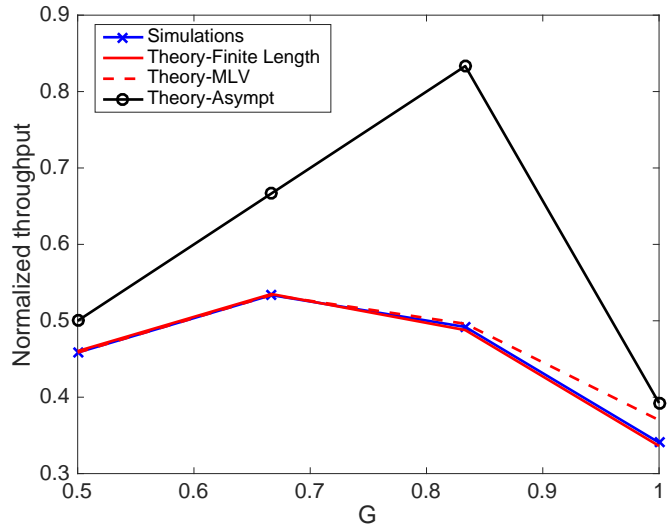}
\caption{Comparison of the theoretical and simulation results for different $(k,t)$ pairs in the case of $t=6$ and $\Lambda(x)=0.2x + 0.5x^2+0.3x^4$.}   \label{fig:failure_Prob}
\end{center}
\end{figure}

It is worth noting that the cardinality of $\mathcal{D}$ and $\mathcal{N}$ both scales with   $k$ and $t$.   However, the probability of experiencing a given reduced vector and a given degree vector can be easily derived from \eqref{fig:failure_Prob}. Therefore,  an approximated PLR can be   evaluated  by  performing the iterative procedure  $Q_u(k,  \mathbf{n})$   only for the most likely  reduced vectors. This substantially  reduces the computational complexity while preserving accuracy.


\section{Numerical Results}
\label{sec:results}
We now provide the simulation results   to validate    the proposed solution in finite-length systems, i.e., with small size MAC frames $t\in[4,7]$. We consider different settings with $k$ sources and $t$ time slots. 
 For each $(k,t)$ pair we consider  different transmission probabilities, i.e., different degree distributions $\Lambda(x)$,  following \cite{Liva:J11}.  For each of these scenarios, we evaluate the decoding error probability from \eqref{eq:plr2_B}. Then,   for each   $\Lambda(x)$, we generate $1000$ realizations of collision matrices and simulate the IRSA protocol and the SIC decoding with belief propagation and we evaluate $u/k$. We then average this ratio over the $1000$ realization to evaluate the average loss probability. 
\begin{table}[t]
\centering
\caption{$k=4$, $t=6$, $\Lambda(x) = 0.25x^2+0.75x^3$.}\label{tab:res1}
\begin{tabular}{c c c}
\toprule
u             & $\Pr [U = u]$ theory  & $\Pr [U = u]$ sim \\\hline
2  &  0.140730 & 0.141390 \\
3  &  0.130158 & 0.130110 \\
4  &  0.094203 & 0.093460 \\
$P_L$ &  0.262186 & 0.261738  \\
\bottomrule
\end{tabular}
\end{table}
\begin{table}[t]
\centering
\caption{ Comparison of the decoding failure probability $P_L$ both from theory or by simulation.}  
\label{tab_comparison}
\begin{tabular}{c|c|c|c|c|c}
\toprule
 $G$     &  Simulation    &  \cite{Liva:J11} & \cite{Paolini:C15} &   \eqref{eq:plr2_B}   &      MLV   \\
\hline
 0.5     &    0.13    & 0	 & 0.17   & 0.14 & 0.14    \\
  0.67  &    0.34     &  0	 &  0.58  & 0.35 & 0.35 \\
  0.8    &    0.75   &  0.44    &   0.98  &    0.74 &    0.77 \\
\bottomrule
\end{tabular}
\end{table}

  We now provide simulation results in terms of   normalized throughput,  defined as  $(1-P_L)k/t$. This metric is usually adopted to evaluate the performance of   MAC strategies and it directly reflects the error probability $P_L$.  In Fig. \ref{fig:failure_Prob}, we provide the normalized throughput  as a function of the traffic $G=k/t$ for a scenario with $t=6$  and $\Lambda(x)=0.2x+0.5x^2+0.3x^4$. Results are provided for both simulation results and theoretical ones, namely the finite length analysis proposed in this work and the asymptotic analysis derived in \cite{Liva:J11}. 
  We also provide an approximated solution (labeled MLV --- most likely vectors), where the iterative evaluation of $Q_u$ in \eqref{eq:plr2_B} is performed only over the occupancy vector with a probability $P(\mathbf{d}) \geq 10^{-3}$.  
The results show a  weak match  between asymptotic theory and the simulations results, from here the need for finite length analysis.  From the results, we also observe a   good match between finite length theory (both exact and approximated)  and simulations, showing the accuracy of our study. 
The model is validated also in the   results provided from  Table  \ref{tab:res1}, where  we provide the   final packet loss rate $P_L$ but also  a partial performance of the decoding process (i.e.,   the probability of stopping the decoding step at $u$ unknown denoted by $Pr[U=u]$).  The good match between theory and simulation is confirmed  in these experiments.  

Finally, in Table  \ref{tab_comparison} we compare our analysis with the asymptotic analysis of \cite{Liva:J11}  and the finite-length analysis of \cite{Paolini:C15}. We see that, especially for small value of the traffic network $G$, the asymptotic analysis is far away from the actual performance, and that our study is more precise than \cite{Paolini:C15} especially for large values of the traffic network $G$.    This accuracy comes at a price of a large computational complexity. Because of the complexity factor,   \eqref{eq:plr2_B} might be too expensive to evaluate for realistic  MAC frames (hundreds of time slots). However, in Table  \ref{tab_comparison} we observe that   the approximated solution MLV nicely scales with the MAC frame without significantly affecting the accuracy.  
  

\section{Conclusions}
We carried out  an evaluation of the     IRSA performance in finite-length settings, using   combinatorial theory and matrix-occupancy  theory. Simulation results validate the derived analysis for small MAC frames and show the improved match between theory and simulation results with respect to the state of the art performance studies. 

\bibliographystyle{IEEEtran}
\bibliography{LDPC_SlottedAloha}

\end{document}